\documentclass[12pt,preprint]{aastex}
\usepackage{emulateapj5}

\newcommand{\cf}{{\ifmmode{C_f}\else{$C_{f}$}\fi}}
\newcommand{\zem}{{\ifmmode{z_{em}}\else{$z_{em}$}\fi}}
\newcommand{\zabs}{{\ifmmode{z_{abs}}\else{$z_{abs}$}\fi}}
\newcommand{\kms}{{\ifmmode{{\rm km~s}^{-1}}\else{km~s$^{-1}$}\fi}}
\newcommand{\delv}{{\ifmmode{\Delta v}\else{$\Delta v$}\fi}}
\newcommand{\vej}{{\ifmmode{v_{ej}}\else{$v_{ej}$}\fi}}
\newcommand{\cm}{{\ifmmode{{\rm cm}^{-1}}\else{cm$^{-1}$}\fi}}
\newcommand{\cmm}{{\ifmmode{{\rm cm}^{-2}}\else{cm$^{-2}$}\fi}}
\newcommand{\cmmm}{{\ifmmode{{\rm cm}^{-3}}\else{cm$^{-3}$}\fi}}
\newcommand{\lya}{Ly$\alpha$}
\newcommand{\1}{I}
\newcommand{\2}{I\hspace{-.1em}I}
\newcommand{\3}{I\hspace{-.1em}I\hspace{-.1em}I}

\newcounter{species} 
\def\ion#1#2{\setcounter{species}{#2}#1$\;${\scriptsize\Roman{species}}\relax}

\slugcomment{\today} 

\shorttitle{Multi-Sightline Spectroscopy of SDSS~J1029+2623}
\shortauthors{Misawa et al.}

\begin{document}

\title{Spectroscopy along Multiple, Lensed Sightlines through
  Outflowing Winds in the Quasar SDSS~J1029+2623\altaffilmark{1}}

\footnotetext[1]{Based on data collected at Subaru Telescope, which is
  operated by the National Astronomical Observatory of Japan.}

\author{Toru Misawa\altaffilmark{2},
        Naohisa Inada\altaffilmark{3},
        Ken Ohsuga\altaffilmark{4,5},
        Poshak Gandhi\altaffilmark{6}, 
        Rohta Takahashi\altaffilmark{7}, and
        Masamune Oguri\altaffilmark{8}}

\altaffiltext{2}{School of General Education, Shinshu University,
  3-1-1 Asahi, Matsumoto, Nagano 390-8621 Japan}
\altaffiltext{3}{Department of Physics, Nara National College of
  Technology, Yamatokohriyama, Nara 639-1080, Japan}
\altaffiltext{4}{National Astronomical Observatory of Japan, Osawa,
  Mitaka, Tokyo 181-8588, Japan}
\altaffiltext{5}{School of Physical Sciences, Graduate University of
  Advanced Study (SOKENDAI), Shonan Village, Hayama, Kanagawa
  240-0193, Japan}
\altaffiltext{6}{Institute of Space and Astronautical Science (ISAS),
  Japan Aerospace Exploration Agency, 3-1-1 Yoshinodai, chuo-ku,
  Sagamihara, Kanagawa 252-5210, Japan}
\altaffiltext{7}{Department of Natural and Physical Sciences,
  Tomakomai National College of Technology, Tomakomai 059-1275, Japan}
\altaffiltext{8}{Kavli Institute for the Physics and Mathematics of
  the Universe (Kavli IPMU, WPI), University of Tokyo, Chiba 277-8583,
  Japan}

\email{misawatr@shinshu-u.ac.jp}

\begin{abstract}
We study the origin of absorption features on the blue side of the
\ion{C}{4} broad emission line of the large-separation lensed quasar
SDSS~J1029+2623 at \zem $\sim$2.197.  The quasar images, produced by a
foreground cluster of galaxies, have a maximum separation angle of
$\theta$ $\sim$ 22$^{\prime\prime}\!\!$.5. The large angular
separation suggests that the sight-lines to the quasar central source
can go through different regions of outflowing winds from the
accretion disk of the quasar, providing a unique opportunity to study
the structure of outflows from the accretion disk, a key ingredient
for the evolution of quasars as well as for galaxy formation and
evolution. Based on medium- and high-resolution spectroscopy of the
two brightest images conducted at the Subaru telescope, we find that
each image has different intrinsic levels of absorptions, which can be
attributed either to variability of absorption features over the time
delay between the lensed images, $\Delta t$ $\sim$ 744~days, or to the
fine structure of quasar outflows probed by the multiple sight-lines
toward the quasar. While both these scenarios are consistent with the
current data, we argue that they can be distinguished with additional
spectroscopic monitoring observations.
\end{abstract}


\section{Introduction \label{sec1}}
Quasars are routinely used as background sources to study the gaseous
phases of intervening objects via absorption-line diagnostics.  These
absorption lines have their origins both in {\it intervening} objects
(i.e., foreground galaxies, inter-galactic medium (IGM), quasar host
galaxies) and in sources that are {\it intrinsic} to the quasars.
One of the most promising candidates for the intrinsic absorbers is an
outflowing wind from the accretion disk of quasar central engines.
The outflow is accelerated by magnetocentrifugal forces
\citep{eve05,dek95}, radiation pressure in lines and continuum
\citep{mur95,pro00}, and/or by thermal pressure force (e.g.,
\citealt{bal93,kro01,che05}).  The outflowing winds play an important
role in three ways: 1) the extraction of angular momenta from disks
allows accretions to proceed (e.g.,
\citealt{bla82,eme92,kon94,eve05}), leading to growth of black holes,
2) the disk outflow also provides energy and momentum feedback to
interstellar media of host galaxies and to intergalactic media (IGM),
and inhibits star formation activity (e.g., \citealt{spr05}), 3)
outflowing winds may induce the metal enrichment of the IGM (e.g.,
\citealt{ham97b,gab06}).  Thus, the physical conditions of the outflow
not only promote the evolution of quasars themselves, but greatly
impact the surrounding environments.

Among absorption lines, broad features (hereafter, BALs; FWHM $\geq$
2000~\kms) are easily identified as being intrinsic, because it is
almost impossible for foreground objects to produce very broad and
smooth line profiles.  BALs are detected in about 10 --
20\%\footnote[1]{However, we may under-estimate the detection rate of
  BALs based on flux-limited optical surveys, because \citet{cha00}
  found that approximately 35\% of radio-quiet gravitational lensed
  quasars contain BAL features.} of optically selected quasars (e.g.,
\citealt{hew03,rei03a}), and their detection rate is slightly higher
in radio-quiet quasars (e.g., \citealt{sto92,bec01,gre06}).  BALs are
thought to originate in the outflowing winds when our sight-line
intersects this component.  This idea is supported by the fact that
there are no significant differences in the properties of quasars with
BALs (BAL~QSOs) and those without BALs (non-BAL~QSOs)
\citep{wey91,rei03a}.  Moreover, quasar spectra tend to be redder when
BAL profiles, especially those with low ionization absorption lines
(i.e., LoBALs and FeLoBALs), are observed, which is probably caused by
dust reddening in the outflows (e.g., \citealt{spr92,yam99}).  Thus,
quasars with and without BALs may intrinsically be a single
population, although the evolutionary phase of quasars could also
affect the detection rate of BALs (e.g., \citealt{voi93}).  The only
difference is whether the sight-line passes through the outflowing
wind or not in this orientation scheme.  Thus, BAL QSOs have
traditionally been the best targets for the study of outflows (e.g.,
\citealt{wey91,bec97,gib09,cap12}).

In addition to BALs, a fraction of narrow absorption lines (hereafter
NALs; FWHM $\leq$ 500~\kms) and mini-BALs (an intermediate category
between BALs and NALs) have also been suggested to be physically
associated with quasars.  However, the origin of NALs and mini-BALs
are still under debate.  Each of the above classes of absorption
features could represent either different lines of sight through the
outflowing wind to the quasar continuum source or different stages in
the evolution of the absorbing gas parcels (e.g.,
\citealt{ham04,mis05}).  The observed fraction of optically-selected
quasars hosting BALs ($\sim$10--20) and NALs ($\sim$20--50\%; e.g.,
\citealt{ves03,wis04,mis07a,nes08}) constrains the solid angle
subtended by the dense portion of the wind to the central engine.
However, such a statistical treatment requires the assumption that all
quasars are identical. This remains an assumption because in past
studies, it has only been possible to trace {\it single} sight-lines
(i.e., probe a one dimension profile) toward the nucleus of each
quasar.

Gravitationally lensed images of background quasars are our only tools
for the study of the three-dimensional geometry of the absorbers
(e.g., \citealt{cro98,rau99,lop05}). Indeed, this technique has
frequently been applied for investigating intervening absorption
lines, produced by galaxies or IGM on scales of up to several
kilo-parsecs. However, these traditional quasar lenses are not useful
for multi-angle studies of the AGNs themselves because their
separation angles ($\theta$ $\sim$ a few arcsec) are too small to
separate internal structures in the vicinity of the quasars. On the
other hand, \citet{ina03} discovered a lensed quasar, SDSS~J1004+4112,
with a maximum image separation of $\theta$ $\sim$
14$^{\prime\prime}\!\!$.6, which is the first example of a quasar
lensed by a cluster of galaxies rather than a single massive
galaxy. If there exists outflow gas at a distance of
$\sim$1~kpc\footnote[2]{However, we usually have only loose
  constraints on the absorber's distance, spanning from $r$ = 0.01 to
  1000~pc (e.g., \citealt{elv00,dek01}).} from the continuum source as
measured for some specific quasars \citep{dek01,ham01}, the large
separation angle of SDSS~J1004+4112 translates into a physical
distance of $\gtrsim$0.1~pc, with which we may trace outflowing winds
with different physical properties unless their transverse sizes are
much larger than sub-parsec scale. Taking advantage of this large
image separation, \citet{gre06} proposed that the differences in
emission lines between the lensed images seen in their follow-up
spectra can be explained by differential absorptions along each
sight-line, although the spectra do not show explicit intrinsic
absorption features.

\begin{figure}
 \begin{center}
  \includegraphics[width=12cm,angle=0]{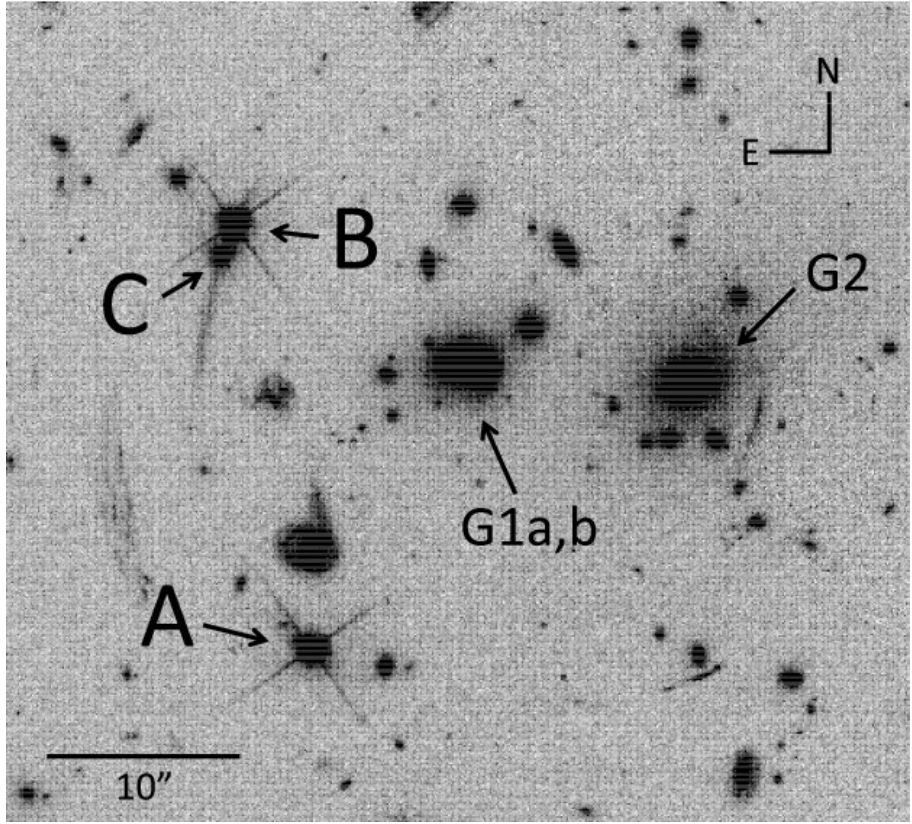}
 \end{center}
 \caption{F814W-band image of SDSS~J1029+2623 taken with HST/ACS
   \citep[GO-12195;][]{ogu12}.  Three objects labeled with A, B, and C
   are lensed images of the quasar at \zem\ = 2.197, while those with
   G1a, G1b, G2 are member galaxies of a foreground lensing cluster at
   $z$ $\sim$ 0.6. Angular separations of the images are
   $\theta_{AB}$ $\sim$ 22$^{\prime\prime}\!\!$.5,
   $\theta_{AC}$ $\sim$ 21$^{\prime\prime}\!\!$.0, and
   $\theta_{BC}$ $\sim$ 1$^{\prime\prime}\!\!$.85.\label{f1}}
\end{figure}

\begin{figure}
 \begin{center}
  \includegraphics[width=12cm,angle=0]{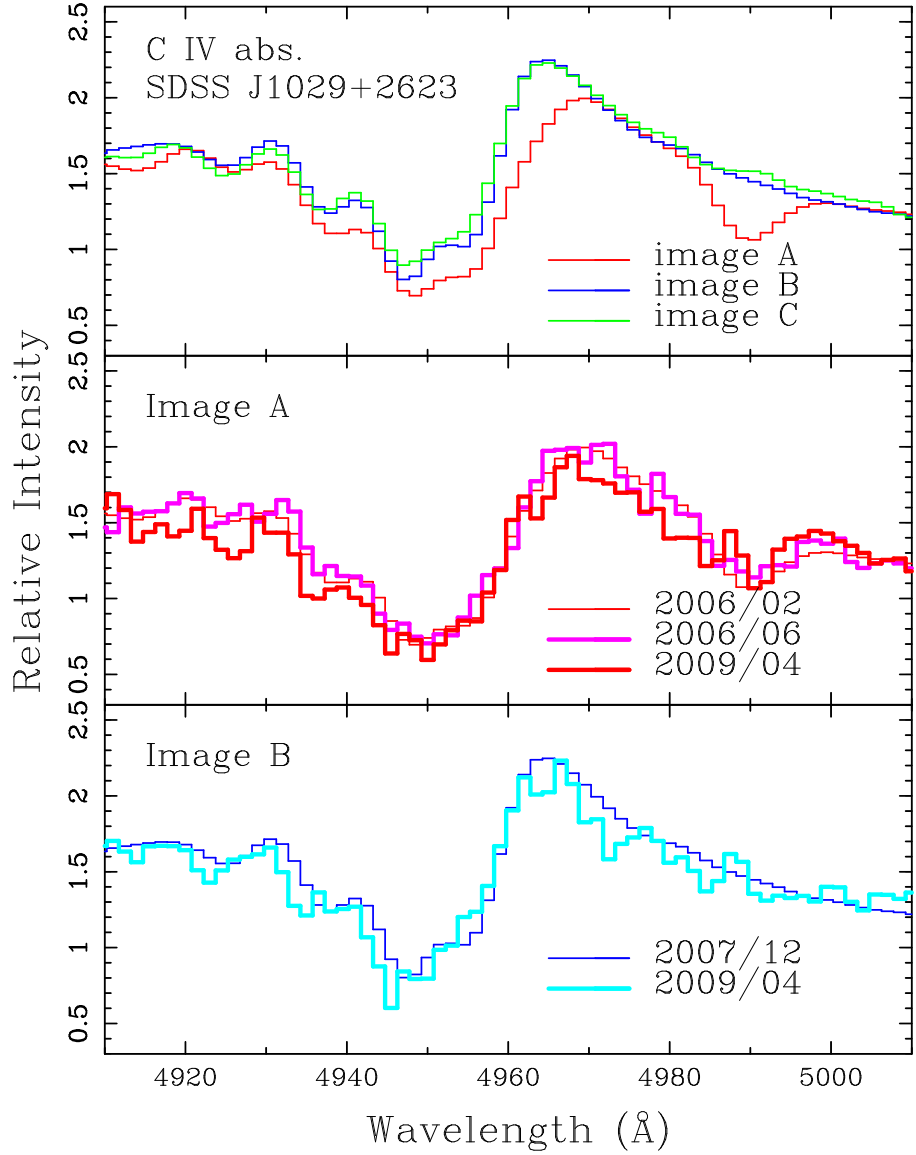}
 \end{center}
 \caption{Medium-resolution spectra ($R$ $\sim$ 500): (Top) Zoomed
   spectra of images~A, B, and C of SDSS~J1029+2623 around the
   associated \ion{C}{4} absorption line \citep{ina06,ogu08}. Fluxes
   are arbitrary and normalized to compare the absorption profiles of
   the three images. An additional absorption line at
   $\sim$4990\AA\ present only in the image~A spectrum is
   \ion{Fe}{2}~$\lambda$2600 at \zabs\ $\sim$ 0.9187. (Middle) Spectra
   of image~A, taken in 2006 twice \citep{ina06,yor00} and in 2009 by
   us.  (Bottom) Same as the above panel, but for image~B, taken in
   2007 \citep{ogu08} and in 2009 by us.\label{f2}}
\end{figure}

The second large-separation lensed quasar SDSS~J1029+2623
\citep{ina06} has an image separation of $\theta$ $\sim$
22$^{\prime\prime}\!\!$.5, and therefore is the largest quasar lens
currently known. \citet{ogu08} identified the third image of this lens
system (see Figure~\ref{f1}). As shown in Figure~\ref{f2}, all the
spectra of quasar images~A, B, and C have absorption features on the
blue side of the \ion{C}{4} emission lines, which implies that each
sight-line passes through the outflowing wind. In what follows, we
call this feature ``the associated absorption line''.\footnote[3]{A
  term {\it associated} absorption line is traditionally used for
  absorption features that fall within 5000\kms\ of quasar emission
  redshift \citep{wey79}. We use this term for this specific
  absorption feature throughout the paper.} Moreover, we see a clear
difference in the absorption features between image~A and
images~B/C. The former has a broader and shallower profile and an
additional redshifted absorption line at $\sim$4990\AA, while the
latter two have narrower and deeper profiles (Figure~\ref{f2}).
After correcting for magnification, the Eddington ratio of the quasar
is estimated to be $L/L_E$ $\sim$ 0.11. This value is consistent with
the typical Eddington ratio of bright quasars, $L/L_E$ $\sim$ 0.07 --
1.6 \citep{net07}.

In this paper, we discuss the origin of these differential absorption
profiles in SDSS~J1029+2623. The likely scenarios include (i)
time-variation of a single absorber covering both sight-lines toward
the quasar between the time-delay of images~A and B/C, which is
$\sim$744 days in the observed frame according to monitoring
observations by \citet{foh12}\footnote[4]{The time-delay between
  images~B and C is almost negligible, $\sim$2--3 days \citep{foh12}.}
(hereafter, {\bf scenario~\1}), (ii) each sight-line penetrating
different absorbers or different regions of a single absorber toward
the quasar as proposed by \citet{gre06} ({\bf scenario~\2}), and (iii)
micro-lensing ({\bf scenario~\3}). Among these, the last scenario has
been already rejected because each image shows common ratios between
radio, optical, and X-ray fluxes \citep{ota12,ogu12}, which is not
expected for micro-lensing. Thus we concentrate on the viability of
scenarios~\1 and \2, based on our medium- and high-resolution
spectroscopic observations of SDSS~J1029+2623.

The structure of this paper is as follows.  In \S2, we describe the
observations and data reduction. We present results in \S3, and
discuss the results in \S 4. We summarize our results in \S 5. We
adopt $\zem = 2.197$ as the emission redshift of the quasar, which was
estimated using broad UV emission lines by \citet{ina06}.  Time
intervals between observations are given in the observed frame
throughout the paper, unless otherwise noted.

\section{Observations and Data Reduction \label{sec2}}
We conducted spectroscopic observation of the two brightest lensed
images\footnote[5]{The observed flux ratio of three lensed-images is A
  : B : C $\sim$ 0.95 : 1.00 : 0.24 in the I-band \citep{ogu08}.}
(i.e., images~A and B) of SDSS~J1029+2623 with Subaru/FOCAS on 2009
April 4, using a similar resolution power to previous observations
($R$ $\sim$ 500) taken about 2--3 years before. This time separation
is comparable to the measured time delay between image~A and
images~B/C \citep[$\sim$744~days;][]{foh12}.  We used the grating with
L600 filter to cover the wavelength range of $\lambda$ = 3700 --
6000\AA. We also used the $0^{\prime\prime}\!\!.4$ slit without pixel
sampling ($\sim$1.34\AA\ per pixel), which enables us to compare our
spectra with the previous observations from the literature
\citep{ina06,ogu08} including SDSS \citep{yor00} by performing
convolution where needed. The sky conditions were moderate although
affected by thin clouds.  The total integration time was 1200~s for
each image, and the data quality of the final spectra are S/N
$\sim$8.5~pix$^{-1}$ and $\sim$7.0~pix$^{-1}$ for images~A and B,
respectively.

We also obtained high-resolution spectra of images~A and B with
Subaru/HDS on 2010 February 10 to resolve the associated \ion{C}{4}
absorption line completely and extract physical parameters such as
column density, line width, ejection velocity from the quasar, and the
covering factor (the fraction of flux from the background source that
is geometrically covered by the foreground absorber) using Voigt
profile fitting.  We used the standard setup, Std-Bc, with a
$1.\!\!^{\prime\prime}2$ slit width ($R$=30,000), covering the
wavelength range of 3390--4210 \AA\ on the blue CCD and
4280--5110~\AA\ on the red CCD.  This configuration covers \lya,
\ion{N}{5}, \ion{Si}{4} as well as the \ion{C}{4} absorption line at
\zabs\ $\sim$ \zem. The CCD is binned every 4 pixels in both spatial
and dispersion directions (i.e., $\sim$0.05\AA\ per pixel and each
resolution element contains three pixels around 4500\AA). The total
integration times were 14400~s and 14200~s for images~A and B,
respectively, and the final S/N ratios are about 13~pix$^{-1}$ for
both the images.

We reduced both the FOCAS and HDS data in a standard manner with the
software IRAF\footnote[6]{IRAF is distributed by the National Optical
  Astronomy Observatories, which are operated by the Association of
  Universities for Research in Astronomy, Inc., under cooperative
  agreement with the National Science Foundation.}.  As for the HDS
spectra, we divided them by a FLAT frame before normalization in order
to remove the instrumental blaze pattern.  Wavelength calibration was
performed using a Th-Ar lamp.  The observation log is summarized in
Table~\ref{t1}.

\section{RESULTS \label{sec3}}
\subsection{Low resolution spectroscopy \label{sec3.1}}
We examined the suite of Subaru/FOCAS spectra to search for
variability in the \ion{C}{4} lines. If absorption in both the images
show time-variations, it would imply that the profiles are variable
and that the scenario~\1 is more favorable. If this is the case, we
have monitored a single absorber in five epochs on 2006 February 28 +
744~days, 2006 June 29 + 744~days, 2009 April 4 + 744~days (toward
image~A), and on 2007 December 15, 2009 April 4 (toward image~B). If
neither images show time variation, the scenario~\2 will instead be
favored.

\subsubsection{image~A \label{sec3.1.1}}
The associated \ion{C}{4} absorption line of the image~A was
spectroscopically observed three times on 2006 February 28
\citep{yor00}, on 2006 June 29 \citep{ina06}, and on 2009 April 4
(this paper), as shown in Figure~\ref{f2}.  The first one was part of
the SDSS, while the others were taken with Subaru/FOCAS.  For
inter-comparison, we adjust the spectral resolution of all
observations to that lowest available (i.e., $R$ $\sim$ 500) by
resampling and convolution.  The \ion{C}{4} emission and absorption
features show a so-called P-Cygni profile, which implies that the
absorption feature is physically associated to the quasar itself.  An
additional weak absorption feature is seen on the red side of the
\ion{C}{4} emission line at $\lambda$ $\sim$ 4990\AA. However, this is
not \ion{C}{4}, but \ion{Fe}{2}~$\lambda$2600 at \zabs\ $\sim$ 0.9187,
because there exists a corresponding \ion{Mg}{2} doublet at the same
redshift.  We do not see any substantial variation of the \ion{C}{4}
absorption profile (see Figure~\ref{f2}).

\subsubsection{image~B \label{sec3.1.2}}
The associated \ion{C}{4} absorption line of image~B was
spectroscopically observed only twice: once on 2007 December 15 with
Keck/LRIS by \citet{ogu08}, and once on 2009 April 4 by us with
Subaru/FOCAS. \citet{ogu08} also obtained a spectrum of image~C.
Images~B and C show profiles that are almost identical near the
\ion{C}{4} absorption/emission lines, which is probably because the
separation angle between them, $\theta_{BC}$ $\sim$
1$^{\prime\prime}\!\!$.85, is much smaller than the separations
between images~A and B ($\theta_{AB}$ $\sim$
22$^{\prime\prime}\!\!$.5) and images~A and C ($\theta_{AC}$ $\sim$
21$^{\prime\prime}\!\!$.0). Our FOCAS spectrum of image~B is
consistent with those of images~B and C taken about 15 months
($\sim$4.7 month in the rest-frame) before.

Under scenario~\1, time-variation is to be expected.  Instead, we
confirm that the absorption profile of each image is stable (see
Figure~\ref{f2}), suggesting that scenario~\1 is less likely to be
true. However, this conclusion based on our medium-resoluton
spectroscopy will have to be modified significantly, following
investigation of the high-resolution spectra, as we will see in the
next subsection.

\subsection{High resolution spectroscopy \label{sec3.2}}
\begin{figure}
 \begin{center}
  \includegraphics[width=12cm,angle=0]{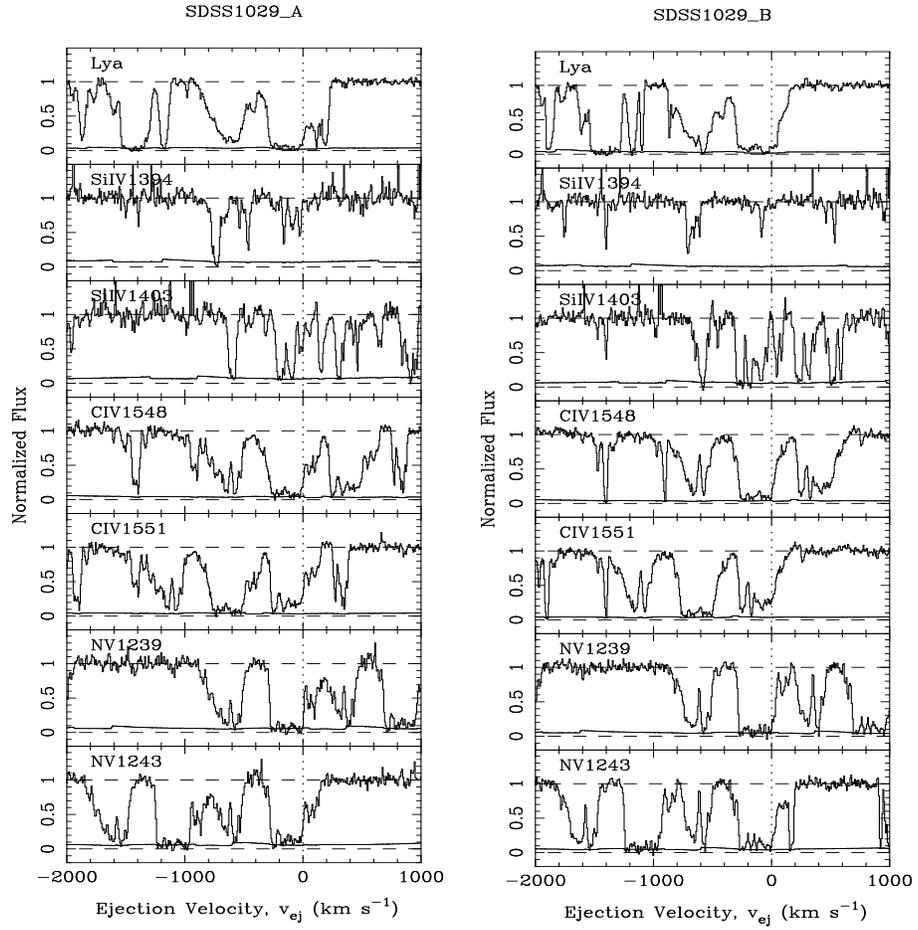}
 \end{center}
 \vspace{1cm}
 \caption{Velocity plot of the associated absorption system toward
   image~A (left) and image~B (right) with high-resolution
   spectra. Vertical dotted lines correspond to the emission redshift
   of the quasar, $z_{em}$ = 2.197.\label{f3}}
\end{figure}

From our high-resolution ($R$ = 30,000) spectra of images~A and B, we
find that the associated \ion{C}{4} absorption line consists of
multiple narrower components (Figure~\ref{f3}). Thus, the system is
not a BAL (or a mini-BAL) but NAL, which means that the features are
not necessarily arising at the outflow.  If the corresponding absorber
is the ISM of the host galaxy, IGM, or a foreground galaxy (i.e.,
intervening absorption), both the differential absorption profiles and
the lack of time-variation toward images~A and B are reasonable
because their sight lines pass through completely different regions
separated by $\sim$kpc or $\sim$Mpc scales with respect to each other.

In order to classify the associated \ion{C}{4} absorption feature as
an intrinsic or intervening feature, we performed a Voigt profile
fitting to measure line parameters.  We used the line fitting software
package {\sc minfit} \citep{chu97,chu03}, with which we can fit
absorption profiles using redshift ($z$), column density ($\log N$ in
\cmm), Doppler parameter ($b$ in \kms), and covering factor (\cf,
defined later in \S 3.2.1) as free parameters.  The revised version of
the code can be applied to self-blending features\footnote[7]{Blue and
  red members of doublets such as \ion{C}{4}, \ion{N}{5}, and
  \ion{Si}{4} are blended with each other if their line widths are
  greater than the velocity distance between the two members (e.g.,
  $\Delta v$ $\sim$ 500 \kms\ in the case of \ion{C}{4} doublet).},
because it fits profiles to the blue and red members of doublets {\it
  simultaneously}, by multiplying the contributions from the two
doublet members \citep{mis07b}.

The fitting results are summarized in Table~\ref{t2}. Column (1) is a
line identification (ID) number. Columns (2) and (3) give the
absorption redshift and the ejection velocity (negative corresponds to
blueshifted) from the quasar emission redshift,
\zem\ $\sim$2.197. Columns (4) and (5) are the column density and its
1$\sigma$ error, columns (6) and (7), the Doppler parameter and its
1$\sigma$ error, and columns (8), (9), and (10), the covering factors
and its upper/lower 1$\sigma$ error. We also applied line fitting to
the \ion{N}{5} doublet. The \ion{Si}{4} doublet cannot be fitted in
both the images, because the red member of the doublet,
\ion{Si}{4}~$\lambda$1403, is severely affected by line contamination.

\subsubsection{Covering Factor Analysis \label{sec3.2.1}}
Covering factor, \cf, is the fraction of coverage of the absorber over
the continuum source and the broad emission line region (BELR) along
the sight-line.  The covering factor can be systematically evaluated
in an unbiased manner by considering the optical depth ratio of
resonant, rest-frame UV doublets of Lithium-like species (e.g.,
\ion{C}{4}, \ion{N}{5}, and \ion{Si}{4}), using the equation
\begin{equation}
C_{f} = \frac{(R_{r}-1)^{2}}{1+R_{b}-2R_{r}}\; ,
\label{eqn:1}
\end{equation}
where $R_{b}$ and $R_{r}$ are the residual (i.e., unabsorbed) fluxes
of the blue and red members of the doublets in the continuum
normalized spectrum (e.g., \citealt{ham97b,bar97,cre99}).  A
\cf\ value less than unity indicates that a portion of the background
source is not occulted by the absorber. This, in turn, means that the
doublet is probably produced by an intrinsic absorber because
intervening absorbers like IGM, ISM, and foreground galaxies usually
have physical scales with a few kilo-parsecs much larger than the
background sources (e.g., \citealt{wam95,bar97}). In general, the
continuum source and the BELR have different covering factors, as
discussed in \citet{gan99}.  The derived coverage fractions can be
significantly affected by uncertainty of the continuum level,
providing unphysical values (i.e., \cf\ $<$ 0 or 1 $<$ \cf),
especially for very weak components whose \cf\ values are close to
1. Therefore, if {\sc minfit} gives unphysical values of \cf\ for some
components in the first trial, we rerun the fit assuming \cf = 1,
following the procedure detailed in \citet{mis07a}.  The fitting
results for \ion{C}{4} and \ion{N}{5} doublets are shown in
Figures~\ref{f4} and \ref{f5}, and fitting parameters are summarized
in Table~\ref{t2}.

\subsubsection{image~A \label{sec3.2.2}}
In our high-resolution spectrum, the associated \ion{C}{4} absorption
line is deblended into at least 16 narrow components at \vej\ $\sim$
$-$1600~\kms\ -- 94~\kms\ from the quasar emission redshift at
\zem\ $\sim$ 2.197, as shown in the \ion{C}{4}~$\lambda$1548 panel of
Figures~\ref{f3} and \ref{f4}.  Among the 16 \ion{C}{4} components,
all have corresponding \lya\ absorption, while only components at
\vej\ $\geq$ $-$846~\kms have corresponding \ion{N}{5} doublets.
Those at \vej\ $\sim$ $-$1400~\kms\ have only subtle \ion{N}{5}
doublets.  It is difficult to tell whether \ion{Si}{4} doublets
corresponding to the \ion{C}{4} system exist or not because the
\ion{Si}{4}~$\lambda$1394 blends with \ion{Si}{2}~$\lambda$1527 at
\zabs\ = 1.9115 and 1.9141, and because the \ion{Si}{4}~$\lambda$1403
is also severely contaminated by \ion{C}{4} doublets at \zabs\ =
1.8909 -- 1.9141.

The absorption component at \vej\ $\sim$ 800~\kms\ in the
\ion{C}{4}~$\lambda$1548 panel is not \ion{C}{4} but instead
\ion{Fe}{2}~$\lambda$2344 at \zabs\ $\sim$ 0.9187.  We also confirm
that a strong absorption feature at $\lambda$ $\sim$ 4900\AA\ only
seen in image~A of Figure~\ref{f2} is \ion{Fe}{2}~$\lambda$2600 at the
same redshift.  It follows that the differential absorption profile of
the associated \ion{C}{4} absorption line between images~A and B seen
in low-resolution spectra (Figure~\ref{f2}) is mainly due to these
physically unrelated \ion{Fe}{2} lines.

\subsubsection{image~B \label{sec3.2.3}}
The associated \ion{C}{4} line is deblended into 12 narrow components
at \vej\ $\sim$ $-$1467 -- 9.4~\kms\ from the quasar emission redshift
as shown in the \ion{C}{4}~$\lambda$1548 panel of Figures~\ref{f3} and
\ref{f4}.  Their general profile is similar to that of the image~A.
Among the 12 components, all have corresponding \lya, while only those
at \vej\ $>$ $-$789~\kms\ have corresponding \ion{N}{5} doublets. On
the other hand, only two \ion{C}{4} components at \vej\ $\sim$
$-$1467~\kms\ and $-$1401~\kms\ have \ion{Si}{4} doublets. The
ionization condition of the components at \vej\ $>$ $-$789~\kms\ is
clearly higher than those at \vej\ $\sim$ $-$1467~\kms\ and
$-$1401~\kms.

The \ion{Si}{4}~$\lambda$1403 lines blend with a clustering of
\ion{C}{4} lines at \zabs\ $\sim$ 1.90.  Absorption features at
\vej\ $\sim$ $-$700~\kms\ in the \ion{Si}{4}~$\lambda$1394 panel and
at \vej\ $\sim$ 200~\kms\ in the \ion{N}{5}~$\lambda$1243 panel are
\ion{Si}{2}~$\lambda$1527 at \zabs\ = 1.9118 and
\ion{C}{2}~$\lambda$1335 at \zabs\ = 1.9788, respectively.

\subsubsection{Other Lines \label{sec3.2.4}}
Because our spectra cover a wide range of wavelength, we also identify
all \ion{C}{4}, \ion{Si}{4}, and \ion{N}{5} doublets from 3800\AA\ --
4210\AA\ on blue CCD chip and from 4280\AA\ -- 5110\AA\ on red CCD
chip, avoiding some unusable regions (4155\AA\ -- 4160\AA\ and
4620\AA\ -- 4740\AA), \lya\ forest, and bad CCD columns.  Toward
image~A, we identify 11 \ion{C}{4} doublets at \zabs\ = 1.6151,
1.8909, 1.8956, 1.8983, 1.9115, 1.9133, 1.9141, 1.9322, 2.1084,
2.1270, 2.1350 and 2 \ion{Si}{4} doublets at \zabs\ = 1.9115 and
1.9138 with $\geq$ 5$\sigma$ detection in addition to the associated
system.  We also find one \ion{Fe}{2} system identified with
\ion{Fe}{2}~$\lambda$2344, \ion{Fe}{2}~$\lambda$2374,
\ion{Fe}{2}~$\lambda$2383, \ion{Fe}{2}~$\lambda$2587,
\ion{Fe}{2}~$\lambda$2600 at \zabs\ $\sim$ 0.9187 as described
above. Toward image~B, we identified 15 \ion{C}{4} doublets at
\zabs\ = 1.6148, 1.6556, 1.6925, 1.7065, 1.8910, 1.8950, 1.8972,
1.9018, 1.9118, 1.9322, 1.9788, 2.1078, 2.1085, 2.1210, 2.1284 and 4
\ion{Si}{4} doublets at \zabs\ = 1.8941, 1.8949, 1.8975, 1.9018 with
$\geq$ 5$\sigma$ detection in addition to the associated system.

\section{DISCUSSION \label{sec4}}
In this section, we discuss the origin (i.e., intrinsic or
intervening) of the associated absorption lines (\S\ref{sec4.1}),
possible geometries toward images~A and B (\S\ref{sec4.2}), and models
of single/multiple sight-line(s) (\S\ref{sec4.3}), respectively.

\subsection{Intrinsic or Intervening \label{sec4.1}}
While BALs have high probability of being physically associated to the
quasars, NALs are difficult to classify as intrinsic or intervening.
With our high-resolution spectra, the associated \ion{C}{4} absorption
line is deblended into multiple narrow components. As for \ion{C}{4}
and \ion{N}{5} doublets (whose absorption features are clearly
detected without being affected by any data defects or line blending),
we measure their physical parameters by performing Voigt profile
fitting with {\sc minfit}. We have several observational results that
support an intrinsic origin for these features, itemized below in
order of relevance.

\begin{figure}
 \begin{center}
  \includegraphics[width=8cm,angle=0]{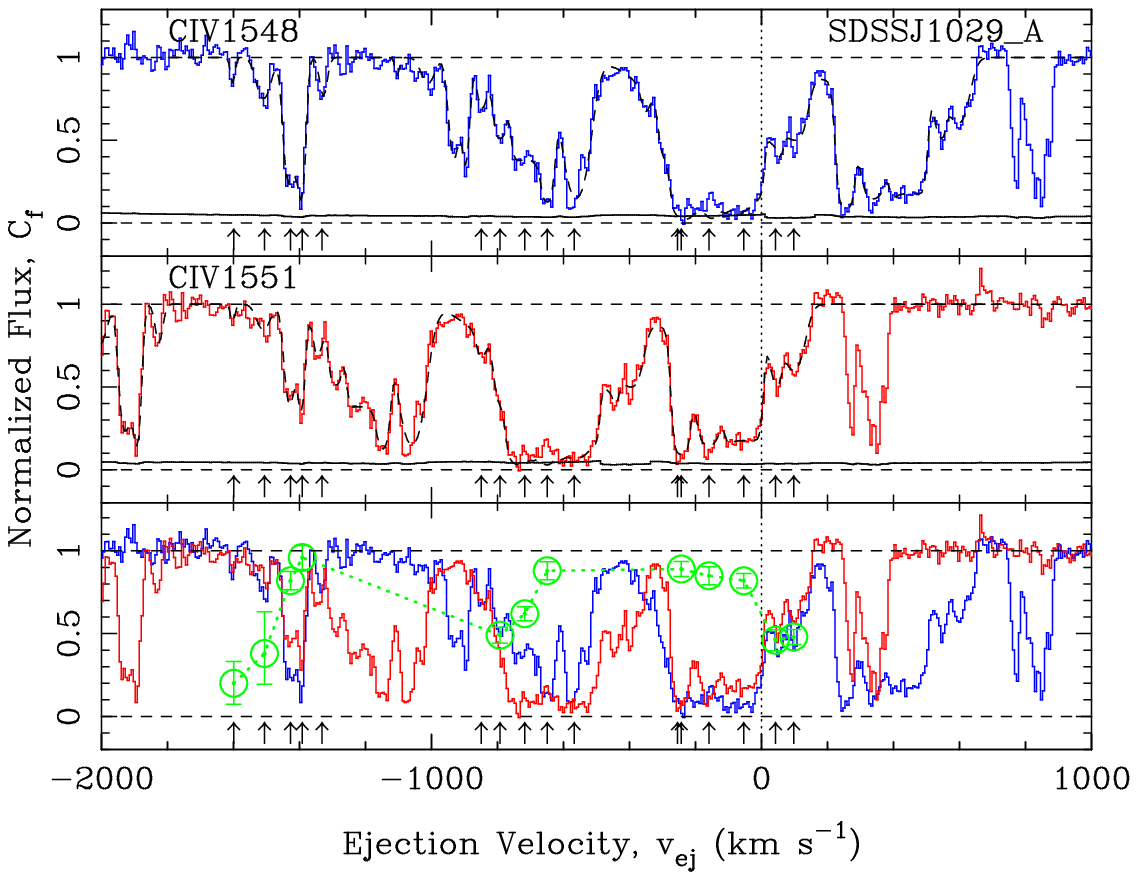}
  \hspace{0.1cm}
  \includegraphics[width=8cm,angle=0]{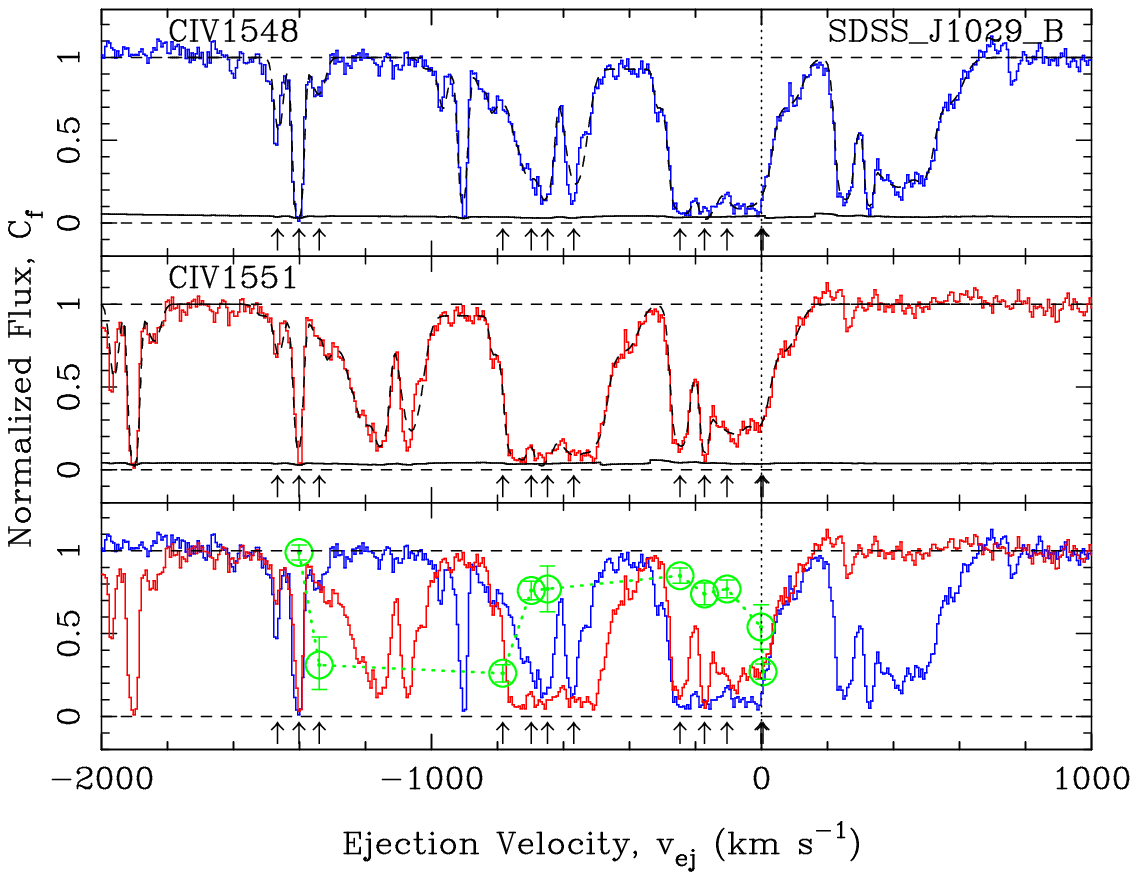}
 \end{center}
 \caption{High-resolution spectra of the images~A (left) and B (right)
   around the associated \ion{C}{4} absorption line. Because
   absorption components are distributed over more than
   \delv\ $\sim$500~\kms, it is self-blended.  The horizontal axis
   denotes relative velocity from the quasar emission redshift at
   \zem\ = 2.197.  Top and middle panels show the profiles of the blue
   and red members of a doublet, with model profiles output by {\sc
     MINFIT} superposed as dashed lines. The positions of the
   kinematic components are marked with upward arrows at the bottom of
   each panel. The bottom panel shows the two profiles together, along
   with the resulting covering factors and their 1$\sigma$
   uncertainties with green open circles.\label{f4}}
\end{figure}

\begin{figure}
 \begin{center}
  \includegraphics[width=8cm,angle=0]{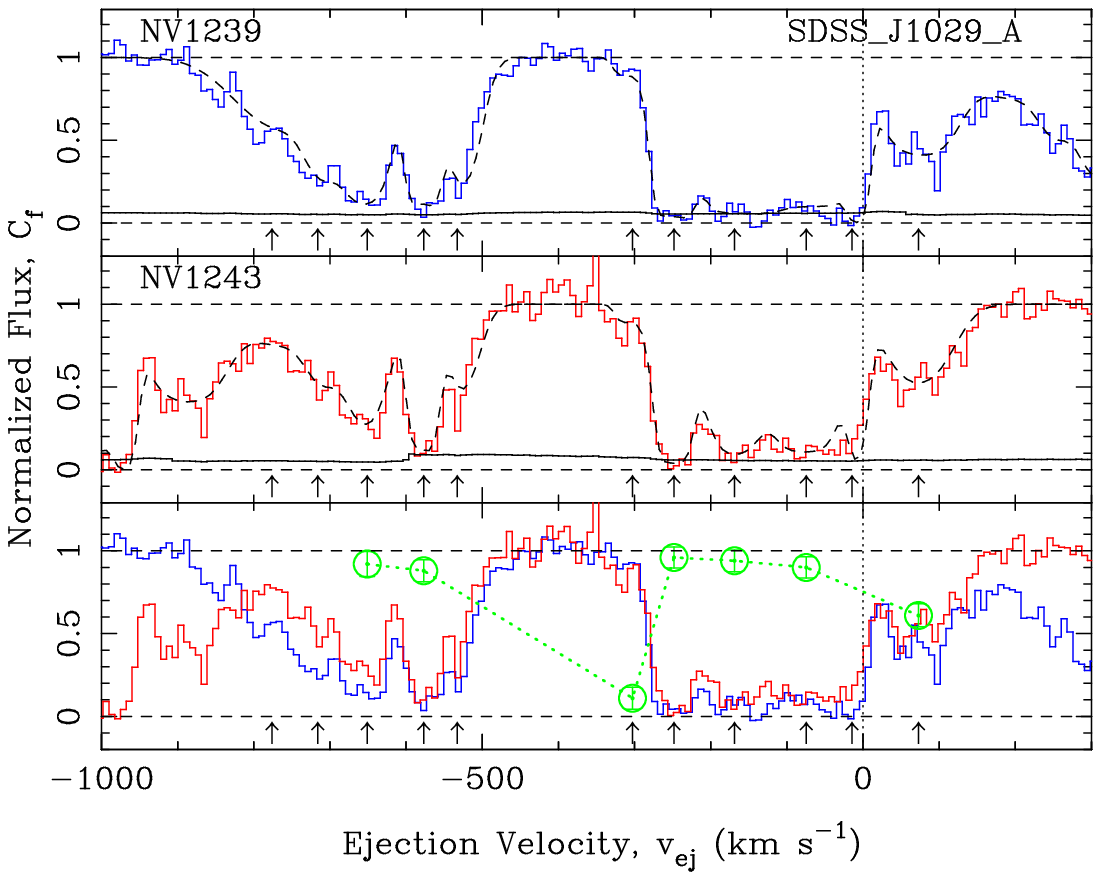}
  \hspace{0.1cm}
  \includegraphics[width=8cm,angle=0]{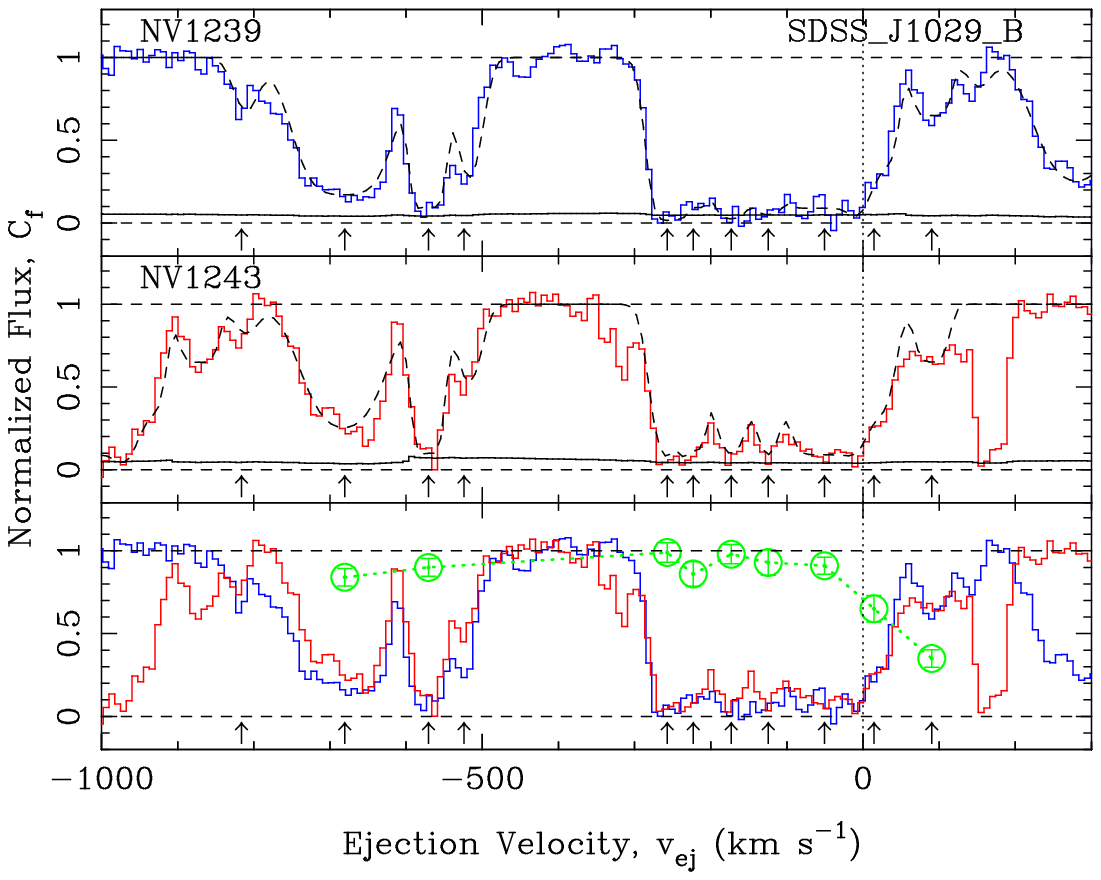}
 \end{center}
 \caption{Same as Figure~\ref{f4}, but for \ion{N}{5}
   doublets.\label{f5}}
\end{figure}

\begin{itemize}
\item[(1)]{The associated \ion{C}{4} NALs show clear evidence of
  partial coverage (see Figure~\ref{f4} and Table~\ref{t2}). Covering
  factors are quite important for ascertaining the location of the
  absorbers.  Among 16 and 12 \ion{C}{4} components in the spectra of
  images~A and B, 6 and 6 components show partial coverage at the
  4$\sigma$ level (i.e., \cf~$+$~4$\sigma$(\cf) $\lessapprox$ 1.0),
  respectively. This supports the physical sizes of the absorbers
  being comparable to or even smaller than the size of the background
  source, suggesting that the system is intrinsic to the quasar itself
  (e.g., \citealt{ham97b,bar97,cre99}).  Interestingly, \ion{N}{5}
  NALs have higher \cf\ values and are consistent with full coverage
  with only a few exceptions.  Covering factors are not necessarily
  identical for all ions from the same absorber, both in BALs (e.g.,
  \citealt{ham97a}) and NALs (e.g., \citealt{mis07a}). In the
  literature, ions in higher ionization states usually have larger
  \cf, as is also the case for the associated system in
  SDSS~J1029+2623.}
\item[(2)]{The associated \ion{C}{4} NALs show the so-called
  line-locking phenomenon (i.e., blue components of \ion{C}{4}
  doublets are aligned with red ones of the following doublets; see
  Figures~\ref{f3} and \ref{f4}). Chance coincidence of such
  alignments is negligibly small.  This is naturally explained by
  radiative acceleration (e.g., \citealt{per78,wey81,fol87}).
  Although several mechanisms have been proposed to explain the
  acceleration of outflowing winds, radiative acceleration almost always
  contributes substantially. Line locking requires that the 
  sight-lines be approximately parallel to the gas motion.}
\item[(3)]{The velocity distribution of the associated \ion{C}{4} and
  \ion{N}{5} NALs shows values beyond 1,000~\kms, which is too large
  for an origin in IGM, foreground galaxy, and ISM of the host galaxy.
  Intervening absorption lines are typically clustered in $\Delta v$
  $<$ 400~\kms\ both for metal lines (e.g.,
  \citealt{sar80,you82,pet94,chu01,pic03}) and for \ion{H}{1} lines
  (e.g., \citealt{lu96,mis04,pen04}). The ISM of most galaxies show
  velocities below this limit.  A distribution with velocities $\geq$
  1,000~\kms\ strongly supports an intrinsic origin due to the outflow
  wind.}
\item[(4)]{These NALs have small ejection velocities from the quasar
  emission redshift, $|\vej|$ $\leq$ 1600~\kms. \citet{wis04} found
  that a high fraction ($\sim$21\%)\footnote[8]{This is a lower limit
    on the fraction since not all systems necessarily show time
    variation.} of the associated NALs within \vej\ = 5,000~\kms\ from
  the quasars, is time-variable (i.e., intrinsic lines).
  \citet{nes08} discovered an excess number density of \ion{C}{4} NALs
  in quasar vicinities, and concluded that almost half of absorbers at
  $|\vej|$ $\leq$ 12000~\kms\ are intrinsic to the quasars, with a
  peak value of $\sim$ 80~\%\ at $|\vej|$ $\sim$ 2000~\kms.}
\end{itemize}

\subsection{Possible Geometries \label{sec4.2}}
Based on multiple medium-resolution spectra, we confirmed that the
associated \ion{C}{4} absorption lines toward both images~A and B
retain their profiles unchanged, while their profiles are clearly
different each other.  This result was originally expected to reject
the time-variation scenario (i.e., scenario~\1).  However, with the
high-resolution spectrograph, the associated \ion{C}{4} absorption
lines are deblended into multiple narrow components, of which two are
foreground absorption lines of \ion{Fe}{2}~$\lambda$2344 and
\ion{Fe}{2}~$\lambda$2600 at \zabs\ $\sim$ 0.9187.  We confirmed that
the differential \ion{C}{4} absorption profiles between the lensed
images seen in medium-resolution spectra are mainly due to these
physically unrelated lines. Thus, we cannot distinguish different
scenarios with medium-resolution spectra alone.

\begin{figure}
 \begin{center}
  \includegraphics[width=12cm,angle=0]{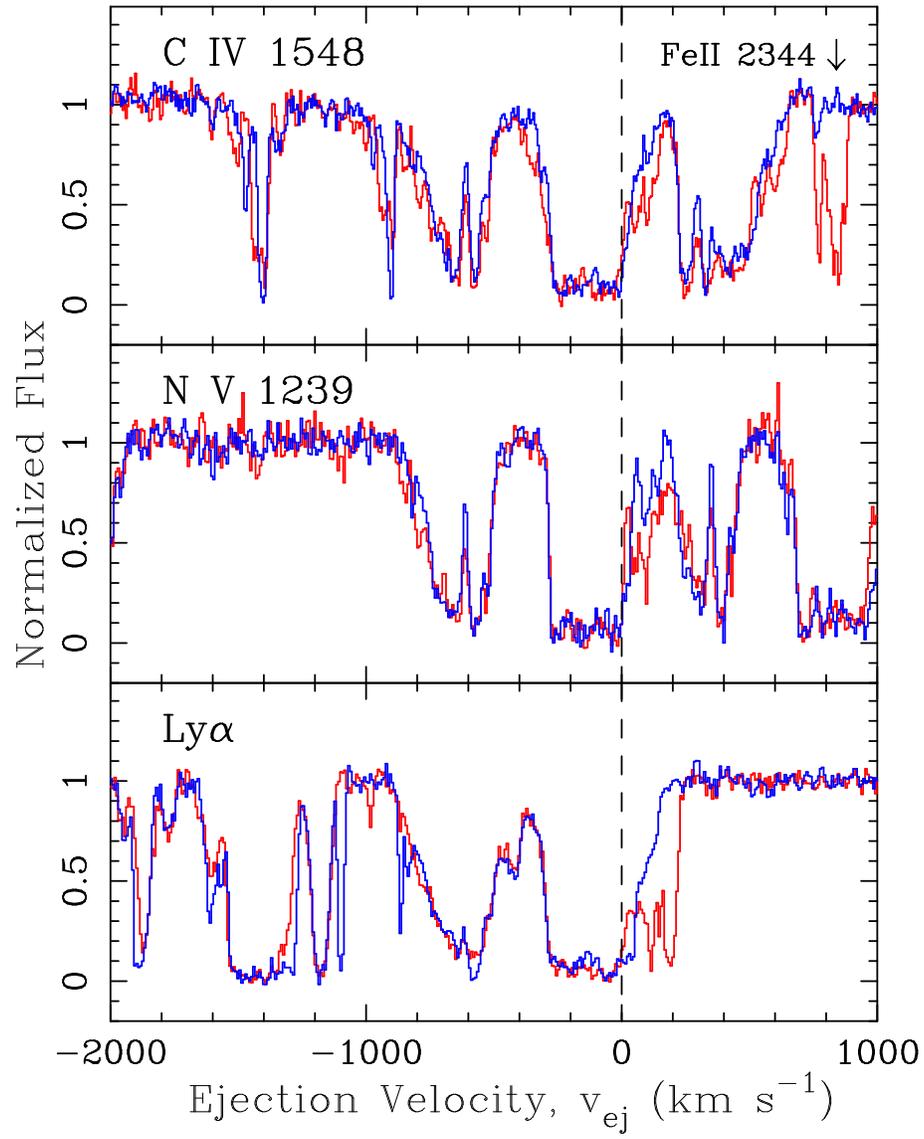}
 \end{center}
 \caption{Comparison of the associated \ion{C}{4}, \ion{N}{5}, and
   \lya\ absorption profiles toward image~A (red) and image~B
   (blue).\label{f6}}
\end{figure}

High-resolution spectra taken with Subaru/HDS provide us with several
important clues regarding the origin of the associated \ion{C}{4}
lines.  First of all, the general profiles of the \ion{C}{4}
absorption lines toward images~A and B are very similar to each other.
This means that the size of the absorber must be larger than the
transverse distance of the two sight-lines.  In the extreme case, the
absorber's location would be very close to the background flux source
so as to make the separation angle almost negligible.
Second, we see a clear difference of absorption profiles between
images~A and B in various structures. For example, at \vej\ $\sim$ 0
-- 200~\kms, there exist additional absorption components in
\ion{C}{4}, \ion{N}{5}, \lya\ panels of Figure~\ref{f6} only in
image~A.  This small difference, discovered only after taking
high-resolution spectra, reminds us of the original question: What is
the source of this difference?
Finally, the most important clue is that a substantial fraction of
absorption components show partial coverage, which means that the
physical sizes of the absorbers are smaller than or comparable to the
background source such as the continuum source and the BELR. While
some intrinsic absorption lines cover only the continuum source (e.g.,
\citealt{ara99}), the covering factors toward the continuum source and
the BELR usually take specific values, depending on the relative
strengths of their fluxes \citep{gan99}. In our case, the residual
flux at $\lambda$ $\sim$ 4948~\AA, at which the \ion{C}{4} emission
line peaks, is almost zero.  This means that absorbers as a whole
cover both the continuum source and the BELR significantly. Therefore,
size estimation of the flux sources is important.

We estimate the size of the BELR ($R_{\rm BELR}$) using the empirical
relation between $R_{\rm BELR}$ and quasar luminosity.  This relation
was originally discovered through reverberation mapping \citep{kas00}
and then extended to brighter quasars with monochromatic luminosities
of $>$10$^{44}$~ergs~s$^{-1}$ to redshifts \zem\ $>$ 0.7 (eq.[A4] in
\citealt{mcl04}).  The monochromatic luminosity of image~A at
$\lambda_r$ = 3000\AA\ in the quasar rest-frame is measured in
\citet{she11} as $\log \lambda L_{3000}$ = 46.21 ergs~s$^{-1}$. After
correcting the magnification factor of image~A, $\mu_A$ = 10.4
\citep{ogu12}, we estimate $R_{\rm BELR}$ to be $\sim$
0.09$^{+0.09}_{-0.05}$ pc, where the main source of 1$\sigma$
uncertainty comes from the scatter in the empirical relation in
\citet{mcl04}.

As to the size of the continuum source ($R_{\rm cont}$), we take five
times the Schwarzschild radius, $5R_S$ = 10$GM_{BH}/c^2$, following
\citet{mis05}, where $R_S$ and $M_{BH}$ are the Schwarzschild radius
and mass of the central black hole.  The virial mass of the central
black hole is already calculated in \citet{she11} using the luminosity
and FWHM of broad emission lines of image~A. Again, after correcting
for magnification, we obtain $\log M_{BH}/M_{\odot}$ = 8.72, and then
$R_{\rm cont}$ is evaluated to be $\sim$ 2.54$\times$$10^{-4}$ pc.
This is almost 300 times smaller than $R_{\rm BELR}$.

\begin{figure}
 \begin{center}
  \includegraphics[width=12cm,angle=0]{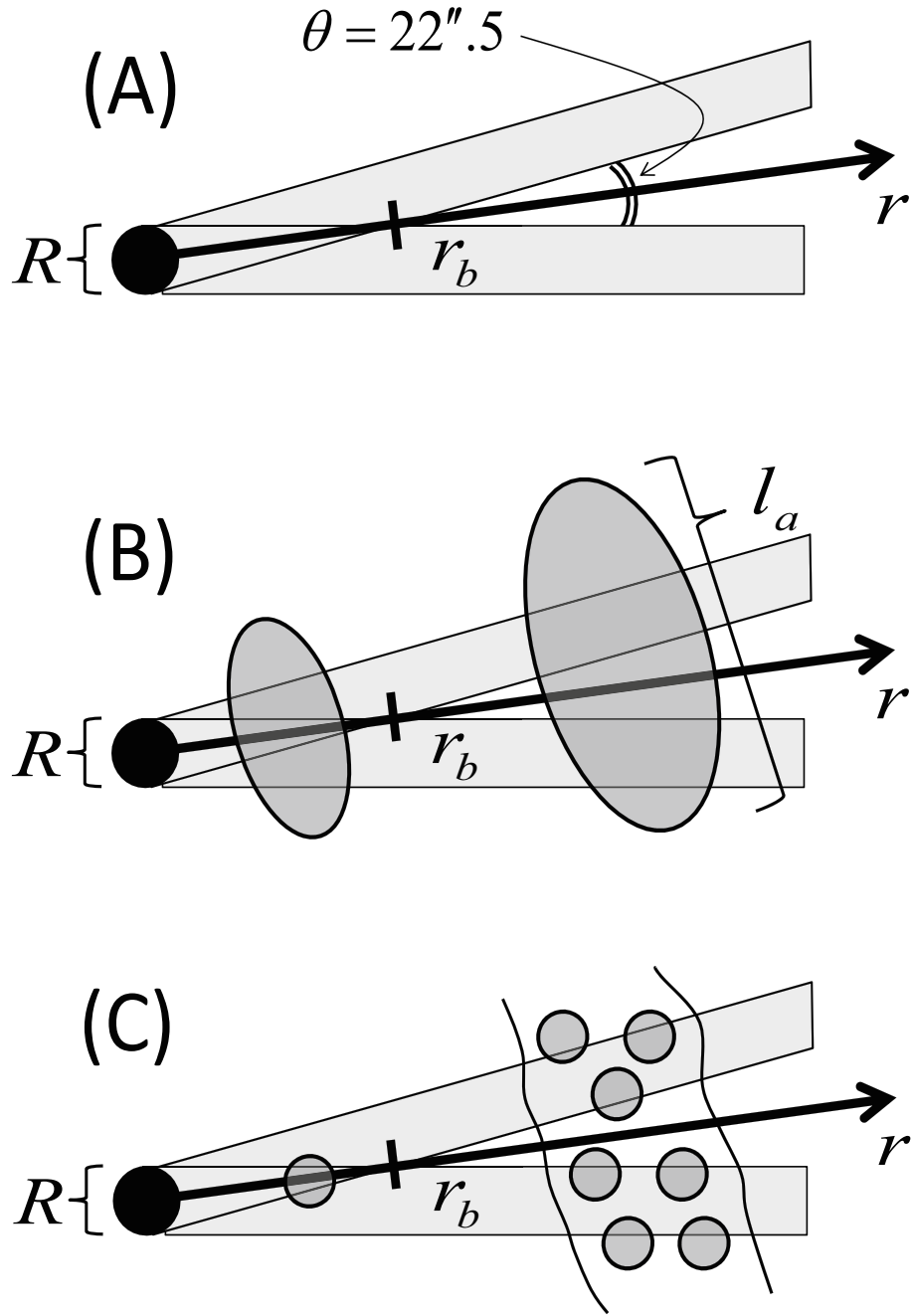}
 \end{center}
 \caption{Possible geometries that could lead to the small
   differential profiles of the associated absorption lines toward the
   two lensed images of SDSS~J1029+2623.  General geometry of two
   sight-lines (A), geometrical models for a single sight-line (B) and
   multiple sight-lines (C) are shown from top to bottom.  A filled
   black circle with a size of $R$ and meshed ovals/circles with a
   size of $l_a$ denote the central flux source and absorbers. The
   radial distance $r_b$ is the boundary distance. A region surrounded
   by two curves in (C) is a filamentary/sheet-like
   structure.\label{f7}}
\end{figure}

The size of the background flux source ($R$) and the absorber's
distance from the central flux source ($r$) decide geometry, i.e., a
single sight-line or multiple sight-lines.  The transverse distance of
the two sight-lines ($a$) can be calculated as $a$ = $r \theta$, where
$\theta$ is the separation angle of the two sight-lines seen from the
flux source (which is very similar to the separation angle seen from
us, $\theta$ $\sim$ 22$^{\prime\prime}\!\!$.5).  At a distance, where
$R$ $\leq$ $a$ (= $r \theta$) is satisfied, the two sight-lines become
fully separated with no overlap.  We call this distance a boundary
distance ($r_b$), hereafter (see Figure~\ref{f7}A).  The boundary
distance is $r_b$ $\sim$2.3~pc if only the continuum source is the
flux source, and $\sim$788 pc if the BELR is also the background
source. The latter is comparable to the distance of outflowing gas
measured by \citet{dek01} and \citet{ham01} for other quasars.
The geometry also depends on the absorber's size ($l_a$).  If the size
of the absorber (whose internal sub-structure is ignored here) is much
larger than the transverse distance (i.e., $l_a$ $\gg$ $r\theta$), a
single sight-line scenario is applicable regardless of the absorber's
distance from the flux source (see Figure~\ref{f7}B).  However, this
geometry cannot be applied to our case because the partial coverage we
found in \S\ref{sec3.2} requires the size of each absorber to be
comparable or smaller than the size of background source.  On the
other hand, if $l_a$ $\leq$ $r\theta$ (Figure~\ref{f7}C), we should
analyze the condition in more detail.
In the case of $l_a$ $\leq$ $r\theta$ $\ll$ $R$ (i.e., the absorber's
distance is smaller than the boundary distance), a substantial
fraction of both sight-lines is covered by a single absorber although
the fraction depends on the distance and the absorber's size.  Small
differences of absorption profiles, seen in the high-resolution
spectra of images~A and B, can be explained either by i) a different
part of the absorber's outskirts covering each sight-line ({\it
  quasi-}scenario~\2), or by ii) time-variation (scenario~\1).
In the case of $l_a$ $\leq$ $R$ $\leq$ $r \theta$ (i.e., the
absorber's distance is larger than the boundary distance), this will
be {\it bona-fide} scenario~\2. One shortcoming of this model is that
we cannot reproduce common absorption profiles seen in the two
sight-lines because the same absorber cannot cover both sight-lines.
We will discuss this problem later.

\subsection{Single sight-line or multiple sight-lines \label{sec4.3}}

\citet{ham11} claimed that NAL absorbers should be located in a low
gravity environment far from the central BH in order to maintain
kinematic stability. Their result suggests that the ejection velocity
of NAL absorbers is larger than the escape velocity so that the
absorber is gravitationally unbound.  If the NAL absorbers in our
target quasar are also in a similar kinematic condition, their radial
distance from the central BH should be $r$ $\geq$ 1.79~pc. This is
close to the boundary distance in the case that only the continuum
source is the background flux source.
As a result, we do not have any conclusive evidence to either accept
or reject scenarios~\1 or \2.  Therefore, we will discuss the absorber's
physical condition further assuming both scenarios in turn below.

\subsubsection{Single sight-line ({\it Scenario~\1}) \label{sec4.3.1}}
If scenario~\1 represents reality, there are two possible origins of
time variability, i) a change of ionization condition of the absorber
(e.g., \citealt{ham11,mis07b}) and ii) the absorber moving across our
sight lines to change the covering factor (e.g.,
\citealt{ham08,gib08}).  Neither situation is applicable for the case
of intervening absorbers as discussed in \citet{nar04}.

If a change in ionization is the origin of variation, we can place
constraints on the electron density and the distance from the flux
source following the procedure by \citet{ham97b} and \citet{nar04}.
We cannot evaluate the specific ionization condition because a wide
range of ionic species (which is necessary for photoionization
modeling) are not detected in our spectra.  Therefore, by adopting the
following assumptions; i) the gas is very close to ionization
equilibrium, ii) the change of ionizing flux is small, and iii)
\ion{C}{4} is the dominant ionization stage of Carbon, we estimate the
electron density to be $n_e$ $\sim$ $1/\alpha \Delta t$ $\sim$
1.78$\times$10$^4$ cm$^{-3}$ by assuming a variation timescale in the
quasar rest frame ($\sim$233 days) as a recombination time ($\Delta
t$), where we use the recombination coefficient of \ion{C}{4}
$\rightarrow$ \ion{C}{3} in gas temperature of 20,000~K
\citep{ham95}. Because the variation timescale is an upper limit of
the recombination time, we should regard the electron density as a
lower limit.
We can also evaluate the distance of the absorber from the flux source
using the prescription of \citet{nar04}.  The ionization parameter
($U$)\footnote[9]{The ionization parameter $U$ is defined as the ratio
  of hydrogen ionizing photon density ($n_{\gamma}$) to the electron
  density ($n_e$), i.e., $U \sim n_{\gamma}/n_e$.} depends on the
bolometric luminosity of the quasar, the continuum shape, distance
from the flux source, and electron density.  By adopting the continuum
shape of \citet{nar04}, we estimate the distance of the absorber to be
$r$ $\leq$0.44 kpc, with an ionization parameter of $U$ = 0.02 (in
which \ion{C}{4} is the dominant ionization stage;
\citealt{ham95,ham97a}), and the bolometric luminosity of the quasar
of $\log L_{\rm bol}$ = 45.87 erg~s$^{-1}$ \citep{she11}. The limits
on the electron density and the absorber's distance are consistent
with the literature (e.g., \citealt{nar04,mis05}).  

If the gas motion (i.e., crossing of our sight-line) is the origin of
variation, we can simply estimate the average crossing velocity
($v_{\rm cross}$ $\sim$ $R/\Delta t$) to be $\sim$390~\kms\ if the
continuum source is the only background source. Again, this is the
minimum velocity because $\Delta t$ is an upper limit on the
variability timescale. The BELR should not be the flux source, because
the corresponding crossing velocity, $\sim$1.32$\times$10$^5$~\kms, is
comparable to the speed of light.  Thus, we conclude that the absorber
covers primarily the UV continuum source with only a small part of the
BELR if the gas motion is the origin of variation.

\subsubsection{Multiple sight-lines ({\it Scenario~\2}) \label{sec4.3.2}}

The main problem of the {\it bona-fide} scenario~\2 model, in which the
absorber's size is smaller than the transverse distance of the two
sight-lines, is that we cannot reproduce closely similar absorption
profiles seen in the two sight-lines because the same absorbers cannot
cover both sight-lines.
A possible solution for this is that there exists a number of clumpy
absorbers or fluctuations in gas density (whose sizes are smaller than
the background source) constituting a filamentary (or a sheet-like)
structure that covers both sight-lines (Figure~\ref{f7}C). Such a
structure is well reproduced above the main body of the outflow by
hydrodynamical simulations (e.g, \citealt{ohs05,pro98,pro99}).

Here, we calculate the radial mass-outflow rate ($\dot{M}$) toward the
two sight-lines summing contributions from all absorption components
as
\begin{equation}
\dot{M} = \sum_i^n\; C_f(i)\; v_{ej}(i)\; R\;
\left(\frac{N_H(i)}{\Delta r(i)}\right) m_p \;,
\label{eqn:2}
\end{equation}
where $N_H$($i$) and $\Delta r$($i$) are the total (\ion{H}{1} $+$
\ion{H}{2}) hydrogen column density and the radial depth of the $i$-th
absorption component toward a sight-line. $R$ and $m_p$ are the size
of background source and the proton mass.
In \S\ref{sec3.2}, we measured ejection velocities of all \ion{C}{4}
and \ion{N}{5} absorption components by applying Voigt profile
fitting.  However, their absolute values could be underestimated
because emission redshifts determined from broad UV emission lines are
systematically blueshifted from the systemic redshift, as measured by
narrow, forbidden lines (see, e.g.,
\citealt{cor90,tyt92,bro94,mar96}).  \citet{tyt92} find a mean
blueshift of the broad UV emission lines relative to the systemic
redshift of about 260~\kms\ and that 90\% of the blueshifts are
between 0 and 650~\kms. We therefore add 260~\kms\ to the \vej\ values
given in Table~\ref{t2} for the calculation herein.

Because our spectra do not detect a wide range of ionic species (which
are necessary for photoionization modeling, as described in
\S\ref{sec4.3.1}), we can measure neither the ionization condition nor
the total hydrogen column density $N$(H). These parameters decide an
absorber's metallicity and radial depth.  Therefore, we simply
calculate $\dot{M}$ assuming all absorption components have the same
ionization condition, metallicity, and radial depth toward both the
sight-lines as follows,
\begin{eqnarray}
  \dot{M} & = & \sum_i^n\; C_f(i)\; v_{ej}(i)\; R\; \left(\frac{N_{ion}(i)
    \times MIC_{{\rm ion}}}{\Delta r(i)}\right) m_p \nonumber\\ &
  \propto & \sum_i^n C_f(i)\; v_{ej}(i)\; N_{ion}(i)\;,
\label{eqn:3}
\end{eqnarray}
where $MIC$ (a metallicity-ionization correction) for \ion{C}{4} and
\ion{N}{5} are defined as follows, respectively,
\begin{equation}
{\rm MIC}_{\; \rm CIV} = \frac{N({\rm H})}{N({\rm C})} \times
\frac{N({\rm C})}{N({\rm C~IV})} = \frac{N({\rm H})}{N({\rm C~IV})}\;,
\label{eqn:4}
\end{equation}
\begin{equation}
{\rm MIC}_{\; \rm NV} = \frac{N({\rm H})}{N({\rm N})} \times
\frac{N({\rm N})}{N({\rm N~V})} = \frac{N({\rm H})}{N({\rm N~V})}\;.
\label{eqn:5}
\end{equation}

Based on the \ion{C}{4} and \ion{N}{5} absorption lines, a difference
of radial mass-outflow rates toward images~A and B is only $\sim$30 --
35\%, which suggests that internal fluctuation of the NAL absorber is
almost negligible within the angular distance of $\theta$
$\sim$ 22$^{\prime\prime}\!\!$.5.
Because $\sim$20 -- 50\%\ of quasars have at least one intrinsic NAL
absorber in their spectra \citep{ves03,wis04,mis07a,nes08}, the global
covering factor of the central flux source surrounded by NAL absorbers
can have values of $\Omega$ $\leq$ 0.8$\pi$ -- 2$\pi$ in terms of
solid angle. Thus, small areas within $\sim$0.002\%\ -- 0.005\%\ of
the total solid angle for NAL absorbers produce similar absorption
profiles with the same order of total column densities.  In other
words, the outflow wind corresponding to the NAL absorbers can be
divided into $\leq$20,000 -- 50,000 small zones with common physical
conditions.

\section{SUMMARY and FUTURE WORK \label{sec5}}
We carried out medium-resolution spectroscopy of the two brightest
images of the quasar SDSS~J1029+2623 to see whether the associated
\ion{C}{4} absorption lines are variable or not, and also
high-resolution spectroscopy in order to measure line parameters and
place strict constraints on the absorber's physical conditions.  Our
main results are as follows:

\begin{enumerate}
\item The associated \ion{C}{4} absorption profile does not show clear
  time-variation toward either sight-lines in medium-resolution
  spectra. With high-resolution spectra, we confirmed that the
  differential line profile is mainly due to the effect of
  \ion{Fe}{2}~$\lambda$2344 and \ion{Fe}{2}~$\lambda$2600 lines which
  arise in a physically unrelated system at \zabs\ $\sim$ 0.9187.

\item In high-resolution spectra, the associated \ion{C}{4} absorption
  line is deblended into more than 10 narrow components, showing i)
  partial coverage, ii) line-locking, iii) large velocity
  distribution, and iv) small ejection velocity. All of these results
  support the associated absorption line being physically related to
  the quasar (i.e., {\it intrinsic} absorption line).

\item The associated absorber probably covers both the continuum
  source and the BELR because the residual flux at $\lambda$ $\sim$
  4948~\AA, at which the \ion{C}{4} emission line peaks, is almost
  zero.  The size of the continuum source and the BELR are evaluated
  to be $R_{\rm cont}$ $\sim$ 2.54$\times$$10^{-4}$ pc and $R_{\rm
    BELR}$ $\sim$ 0.09$^{+0.09}_{-0.05}$ pc, respectively. If the
  associated absorber covers both the flux sources toward our
  sight-lines, their typical scale should be comparable to or smaller
  than $R_{\rm BELR}$ because they show partial coverage.

\item The geometry of the associated absorber depends on the sizes of
  the background flux source ($R$) and the absorber ($l_a$), and the
  absorber's distance from the center ($r$).  If a single sight-line
  model represents reality, the differential line profile impies time
  variation of the absorber. If a change of ionization level is the
  origin of the variation, we can place constraints on the absorber's
  electron density, $n_e$ $\leq$ 1.78$\times$10$^4$ cm$^{-3}$, and the
  absorber's distance from the continuum source, $r$ $\leq$ 0.44~kpc.
  If gas motion is the origin of variation, the continuum source is
  the only background source because the estimated crossing velocity
  is too large (comparable to the speed of light) if the BELR is also
  the flux source.

\item If multiple sight-line model is the real scenario, the following
  condition should be satisfied; the absorber's size ($l_a$) $\leq$
  the size of the background flux source ($R$) $\leq$ transverse
  distance of the two sight-lines ($a$). However, such a condition
  cannot allow a single absorber to cover both sight-lines, in order
  to reproduce similar absorption profiles as seen in the
  high-resolution spectra. A filamentary (or a sheet-like) structure
  made of multiple clumpy gas clouds could solve this issue. Indeed,
  such structures are reproduced by hydro-dynamical simulations.
\end{enumerate}

In order to assess the origin of the associated absorber further, we
need to obtain high-resolution spectra again.  We have shown that
monitoring observations with medium-resolution spectra are not useful
because the associated \ion{C}{4} absorption line in the image~A is
significantly contaminated with physically unrelated foreground
\ion{Fe}{2} absorption lines. Additional high-resolution spectra will
provide us with stronger evidence for scenario~\1 if we see clear
time-variations or for scenario~\2 if the images do not show
variability.

\acknowledgments TM acknowledges support from the Special Postdoctoral
Research Program of RIKEN.  The research was partially supported by
the Japan Society for the Promotion of Science through Grant-in-Aid
for Scientific Research 23740148 and Shinshu University Research Grant
for Exploratory Research by Young Scientists.  This work was also
supported in part by the FIRST program ``Subaru Measurements of Images
and Redshifts (SuMIRe)'', World Premier International Research Center
Initiative (WPI Initiative), MEXT, Japan, and Grant-in-Aid for
Scientific Research from the JSPS (23740161). We also thank Kentaro
Aoki for his comments about data analysis, and Chris Churchill for
providing us with {\sc MINFIT}. We also thank the anonymous referee
for helpful comments and suggestions.

\clearpage


\begin{deluxetable}{cccccccc}
\tabletypesize{\scriptsize}
\setlength{\tabcolsep}{0.04in}
\setcounter{table}{0}
\tablecaption{Log of Observations \label{t1}}
\tablewidth{0pt}
\tablehead{
\colhead{Target}       &
\colhead{Date}         &
\colhead{Order$^a$}    &
\colhead{Instrument}   &
\colhead{$R$}          &
\colhead{exp.}         &
\colhead{S/N$^b$}      &
\colhead{ref.$^c$}     \\
\colhead{}             &
\colhead{}             &
\colhead{}             &
\colhead{}             &
\colhead{}             &
\colhead{(sec)}        & 
\colhead{(pix$^{-1}$)} &
\colhead{}             \\
\colhead{(1)}          &
\colhead{(2)}          &
\colhead{(3)}          &
\colhead{(4)}          &
\colhead{(5)}          &
\colhead{(6)}          &
\colhead{(7)}          &
\colhead{(8)}          
}
\startdata
image~A & 2006 February 28 & 2 & SDSS         &  1800 &       &  9.1 & 1 \\
        & 2006 June     29 & 3 & Subaru/FOCAS &   500 &   600 & 19   & 2 \\
        & 2009 April     4 & 5 & Subaru/FOCAS &   500 &  1200 &  8.5 & 3 \\
image~B & 2007 December 15 & 1 & Keck/LRIS    &  1000 &   800 & 43   & 4 \\
        & 2009 April     4 & 4 & Subaru/FOCAS &   500 &  1200 &  7.0 & 3 \\
image~C & 2007 December 15 & 1 & Keck/LRIS    &  1000 &   800 & 11   & 4 \\
\hline
image~A & 2010 February 10 &  & Subaru/HDS    & 30000 & 14400 & 13   & 3 \\
image~B & 2010 February 10 &  & Subaru/HDS    & 30000 & 14200 & 13   & 3 \\
\enddata
\tablenotetext{a}{Order of observing epoch considering time-delay,
  i.e., adding $\sim$744~days only to the observation of image~A.}
\tablenotetext{b}{Signal to noise ratio around 4700\AA.}
\tablenotetext{c}{References --- (1) Sloan Digital Sky Survey
  \citealt{yor00}; (2) \citealt{ina06}; (3) this paper; (4)
  \citealt{ogu08}.}
\end{deluxetable}

\begin{deluxetable}{cccccccccc}
\tabletypesize{\scriptsize}
\tablecaption{Fitting parameters of C~IV absorbers \label{t2}}
\tablewidth{0pt}
\tablehead{
\colhead{ID}                       &
\colhead{$z_{abs}$$^a$}             &
\colhead{$v_{ej}$$^b$}              &
\colhead{$\log N$}                 &
\colhead{$\sigma$($\log N$)}       &
\colhead{$b$}                      &
\colhead{$\sigma$($b$)}            &
\colhead{$C_f$$^c$}                &
\colhead{$+\sigma$($C_f$)}          &
\colhead{$-\sigma$($C_f$)}          \\
\colhead{}                         &
\colhead{}                         &
\colhead{(\kms)}                   &
\colhead{(\cmm)}                   &
\colhead{(\cmm)}                   &
\colhead{(\kms)}                   &
\colhead{(\kms)}                   &
\colhead{}                         &
\colhead{}                         &
\colhead{}                        \\
\colhead{(1)}                      &
\colhead{(2)}                      &
\colhead{(3)}                      &
\colhead{(4)}                      &
\colhead{(5)}                      &
\colhead{(6)}                      &
\colhead{(7)}                      &
\colhead{(8)}                      &
\colhead{(9)}                      &
\colhead{(10)}                      
}
\startdata
\multicolumn{3}{c}{}  &
\multicolumn{4}{c}{Image~A} &
\multicolumn{3}{c}{}  \\
\cline{4-7} \\
\ion{C}{4} & & & & & & & & \\
 1 & 2.1800 & $-$1599.5 & 13.52 & 0.54 &   6.30 &   2.45 & 0.20 & 0.13 & $-$0.13 \\
 2 & 2.1810 & $-$1505.2 & 13.80 & 0.29 &  25.52 &   3.01 & 0.38 & 0.25 & $-$0.18 \\
 3 & 2.1818 & $-$1429.7 & 14.08 & 0.06 &  18.51 &   1.19 & 0.82 & 0.05 & $-$0.05 \\
 4 & 2.1822 & $-$1392.0 & 13.85 & 0.06 &  13.16 &   0.83 & 0.96 & 0.08 & $-$0.08 \\
 5 & 2.1829 & $-$1326.0 & 12.97 & 0.98 &  14.35 &   37.5 & 1.00$^d$ & ... & ... \\
 6 & 2.1880 &  $-$845.7 & 13.39 & 0.75 &  16.40 &   3.62 & 0.62 & 2.88 & $-$0.88 \\
 7 & 2.1886 &  $-$789.3 & 14.28 & 0.15 &  16.08 &   2.18 & 0.49 & 0.05 & $-$0.05 \\
 8 & 2.1894 &  $-$714.0 & 15.03 & 0.34 &  25.44 &   6.00 & 0.62 & 0.04 & $-$0.04 \\
 9 & 2.1901 &  $-$648.2 & 14.28 & 0.12 &  19.18 &   2.61 & 0.88 & 0.06 & $-$0.06 \\
10 & 2.1910 &  $-$563.6 & 14.28 & 0.09 &  44.19 &   9.87 & 1.00$^d$ & ... & ... \\
11 & 2.1942 &  $-$262.9 & 14.01 & 0.29 & 143.71 &  113.3 & 1.00$^d$ & ... & ... \\
12 & 2.1944 &  $-$244.1 & 14.82 & 0.10 &  21.34 &   1.63 & 0.89 & 0.04 & $-$0.04 \\
13 & 2.1953 &  $-$159.6 & 14.84 & 0.09 &  34.58 &   5.17 & 0.85 & 0.06 & $-$0.06 \\
14 & 2.1964 &   $-$56.3 & 15.65 & 0.20 &  32.62 &   2.83 & 0.82 & 0.04 & $-$0.04 \\
15 & 2.1975 &      46.9 & 14.37 & 0.12 &  16.02 &   2.14 & 0.46 & 0.06 & $-$0.06 \\
16 & 2.1980 &      93.8 & 14.46 & 0.06 &  34.85 &   2.60 & 0.48 & 0.04 & $-$0.04 \\
\\
\ion{N}{5} & & & & & & & & \\
 1 & 2.1889 &  $-$761.1 & 14.18 & 0.60 &  77.80 &   73.1 & 1.00$^d$ & ... & ... \\
 2 & 2.1894 &  $-$714.0 & 13.94 & 0.85 &  27.88 &   31.5 & 1.00$^d$ & ... & ... \\
 3 & 2.1901 &  $-$648.2 & 14.52 & 0.06 &  31.83 &   3.25 & 0.92 & 0.07 & $-$0.07 \\
 4 & 2.1908 &  $-$582.4 & 14.99 & 0.39 &  14.50 &   3.05 & 0.88 & 0.07 & $-$0.07 \\
 5 & 2.1914 &  $-$526.0 & 14.13 & 0.16 &  26.67 &  10.92 & 1.00$^d$ & ... & ... \\
 6 & 2.1938 &  $-$300.4 & 15.25 & 4.07 &  12.93 &  23.71 & 0.11 & 0.07 & $-$0.07 \\
 7 & 2.1943 &  $-$253.5 & 15.03 & 0.20 &  18.61 &   2.16 & 0.96 & 0.06 & $-$0.06 \\
 8 & 2.1952 &  $-$169.0 & 14.92 & 0.11 &  36.29 &   8.38 & 0.94 & 0.07 & $-$0.07 \\
 9 & 2.1962 &   $-$75.1 & 15.11 & 0.15 &  39.00 &  15.02 & 0.90 & 0.06 & $-$0.06 \\
10 & 2.1969 &    $-$9.4 & 14.41 & 0.07 &  12.90 &   1.91 & 1.00$^d$ & ... & ... \\
11 & 2.1978 &      75.1 & 14.69 & 0.08 &  46.82 &   3.58 & 0.61 & 0.07 & $-$0.07 \\
\\
\multicolumn{3}{c}{}  &
\multicolumn{4}{c}{Image~B} &
\multicolumn{3}{c}{}  \\
\cline{4-7} \\
\ion{C}{4} & & & & & & & & \\
 1 & 2.1814 & $-$1467.4 & 13.37 & 0.41 &  16.46 &  18.15 & 1.00$^d$ & ... & ... \\
 2 & 2.1821 & $-$1401.4 & 14.20 & 0.04 &  13.75 &   0.37 & 0.99 & 0.04 & $-$0.04 \\
 3 & 2.1827 & $-$1344.9 & 13.81 & 0.32 &  21.73 &   3.05 & 0.31 & 0.17 & $-$0.15 \\
 4 & 2.1886 &  $-$789.3 & 15.96 & 1.03 &  17.73 &   4.40 & 0.26 & 0.05 & $-$0.05 \\
 5 & 2.1896 &  $-$695.2 & 14.45 & 0.17 &  38.57 &   5.69 & 0.76 & 0.05 & $-$0.05 \\
 6 & 2.1901 &  $-$648.2 & 14.18 & 0.14 &  18.49 &   2.90 & 0.77 & 0.14 & $-$0.14 \\
 7 & 2.1910 &  $-$563.6 & 14.09 & 0.18 &  37.21 &  11.44 & 1.00$^d$ & ... & ... \\
 8 & 2.1944 &  $-$244.1 & 14.67 & 0.07 &  22.64 &   1.35 & 0.85 & 0.04 & $-$0.04 \\
 9 & 2.1952 &  $-$169.0 & 15.51 & 3.21 &   6.63 &   5.29 & 0.74 & 0.06 & $-$0.06 \\
10 & 2.1959 &  $-$103.2 & 15.03 & 0.06 &  81.86 &   6.01 & 0.77 & 0.06 & $-$0.06 \\
11 & 2.1970 &       0.0 & 14.29 & 0.15 &  26.03 &   4.26 & 0.54 & 0.14 & $-$0.14 \\
12 & 2.1971 &       9.4 & 15.52 & 0.23 &  67.71 &   6.19 & 0.27 & 0.04 & $-$0.04 \\
\\
\ion{N}{5} & & & & & & & & \\
 1 & 2.1884 &  $-$808.1 & 13.44 & 0.78 &  20.68 &   37.5 & 1.00$^d$ & ... & ... \\
 2 & 2.1897 &  $-$685.8 & 14.89 & 0.04 &  51.32 &   1.50 & 0.84 & 0.05 & $-$0.05 \\
 3 & 2.1909 &  $-$573.0 & 15.19 & 0.40 &  13.57 &   2.11 & 0.90 & 0.05 & $-$0.05 \\
 4 & 2.1915 &  $-$516.6 & 13.95 & 0.20 &  19.20 &   10.0 & 1.00$^d$ & ... & ... \\
 5 & 2.1943 &  $-$253.5 & 14.58 & 0.13 &  20.22 &   1.92 & 0.99 & 0.06 & $-$0.06 \\
 6 & 2.1946 &  $-$225.3 & 15.30 & 1.85 &   9.69 &   7.26 & 0.86 & 0.07 & $-$0.07 \\
 7 & 2.1952 &  $-$169.0 & 14.65 & 0.07 &  24.92 &   5.53 & 0.98 & 0.06 & $-$0.06 \\
 8 & 2.1957 &  $-$122.0 & 14.54 & 0.17 &  14.51 &   4.85 & 0.93 & 0.08 & $-$0.08 \\
 9 & 2.1965 &   $-$46.9 & 15.34 & 0.29 &  34.20 &   8.34 & 0.91 & 0.05 & $-$0.05 \\
10 & 2.1972 &      18.8 & 14.92 & 0.35 &  20.74 &   3.18 & 0.65 & 0.08 & $-$0.08 \\
11 & 2.1980 &      93.8 & 14.87 & 0.35 &  15.78 &   3.11 & 0.35 & 0.05 & $-$0.05 \\
\enddata
\tablenotetext{a}{Absorption redshift.}
\tablenotetext{b}{Ejection velocity from the quasar emission redshift,
  with negative values denoting blueshifted line.}
\tablenotetext{c}{Covering factor.}
\tablenotetext{d}{Fitting the component, assuming \cf\ = 1.}
\end{deluxetable}
\clearpage


\clearpage

\clearpage

\clearpage

\clearpage

\end{document}